\documentclass{ws-p8-50x6-00}

\newcommand{\ba}[1]{\begin{eqnarray} \label{(#1)}}
\newcommand{\ea}{\end{eqnarray}}

\newcommand{\AmS}{{\protect\the\textfont2
  A\kern-.1667em\lower.5ex\hbox{M}\kern-.125emS}}

\def\be{\begin{equation}}
\def\ee{\end{equation}}
\def\bea{\begin{eqnarray}}
\def\eea{\end{eqnarray}}
\begin{document} 

%\begin{center}
%{\bf 
\title{New Underground Neutrino Observatory --- GENIUS--- in the New Millenium 
	: for Solar Neutrinos, Dark Matter and Double Beta Decay}

\author{H.V. Klapdor-Kleingrothaus}

\address{Max-Planck-Institut f\"ur Kernphysik,\\ 
P.O. Box 10 39 80, D-69029 Heidelberg, Germany\\ 
Spokesman of HEIDELBERG-MOSCOW and GENIUS Collaborations\\
E-mail: klapdor@gustav.mpi-hd.mpg,\\
 Home-page: http://mpi-hd.mpg.de.non\_acc/}
%\end{center}

\maketitle

%%%%%%%%%%%%%%%%%%%%%%%%%% ABSTRACT %%%%%%%%%%%%%%%%%%%%%%%%%%%%%%

\abstracts{
	Double beta decay is indispensable to solve the question of 
	the neutrino mass matrix together with $\nu$ oscillation experiments. 
	The most sensitive experiment since eight years --- 
	the HEIDELBERG-MOSCOW experiment in Gran-Sasso --- already now, 
	with the experimental limit of 
$\langle m_\nu \rangle < 0.26$~eV excludes degenerate $\nu$ mass scenarios 
	 allowing neutrinos as hot dark matter in the universe for the 
	 small angle MSW solution of the solar neutrino problem. 
	 It probes cosmological models including hot dark matter 
	 already now on the level of future satellite 
	 experiments MAP and PLANCK. 
	 It further probes many topics of beyond Standard Model 
%SM 
	 physics at the TeV scale. 
	 Future experiments should give access to the multi-TeV 
	 range and complement on many ways the search for new physics 
	 at future colliders like LHC and NLC. 
	 For neutrino physics GENIUS will allow to test 
	 almost all neutrino mass scenarios allowed by 
	 the present neutrino oscillation experiments. 
	 At the same time GENIUS will cover a wide range of the parameter 
	 space of predictions of SUSY for neutralinos as cold dark matter. 
	 Further it has the potential to be a real-time detector 
	 for low-energy ($pp$ and $^7$Be) solar neutrinos. 
	 A GENIUS Test Facility has just been funded and will 
	 come into operation by end of 2001.}
%%%%%%%%%%%%%%%%%%%%%%%%%% end ABSTRACT %%%%%%%%%%%%%%%%%%%%%%%%%%%%%%

%%%%%%%%%%%%%%%%%%%%%%%%%% Introduction %%%%%%%%%%%%%%%%%%%%%%%%%
\section{Introduction}
	Underground physics can complement in many ways the search for 
	New Physics at future colliders such as LHC and NLC and can serve 
	as important bridge between the physics that will be gleaned 
	from future high energy accelerators on the one and, 
	and satellite experiments such 
	as MAP and PLANCK on the other%
\cite{KK60Y,GEN-prop,KK-NOW00,KK-WEIN98,KK-Bey97,KK-SprTracts00,KK-InJModPh98}.

	The first indication for beyond Standard Model (SM) physics 
	indeed has come from underground experiments 
	(neutrino oscillations from Superkamiokande), 
	and this type of physics will play an even large role in the future.

	Concerning neutrino physics, without double beta decay 
	there will be no solution of the nature of the neutrino 
	(Dirac or Majorana particle) and of the structure of the 
	neutrino mass matrix. 
	Only investigation of $\nu$ oscillations 
	{\em and}\ double beta decay together can lead to an 
	absolute mass scale% 
\cite{KKPS,KKPS-01,KK60Y,KKP-ComNucl,KK-NOON00}.

	Concerning solar neutrino physics, present information on 
	possible $\nu$ oscillations relies on 0.2\% of the solar neutrino flux.
	The total $pp$ neutrino flux has not been measured and also no 
	real-time information is available for the latter.
	Concerning the search for cold dark matter, direct detection of 
	the latter by underground detectors remains 
	indispensable. 
	
	The GENIUS project proposed in 1997% 
\cite{KK-Bey97,GEN-prop,KK60Y,KK-InJModPh98,KK-J-PhysG98} 
	as the first third generation $\beta\beta$ detector, 
	could attack all of these problems with an unprecedented sensitivity. 
	The main goals of GENIUS are dark matter search and double beta 
	decay. 		
	In this paper we shall concentrate on the neutrino physics 
	with emphasis on solar neutrinos, and on some  
	   dark matter aspects. 
	The GENIUS project at this conference is for the first time presented 
	in its capacity of a solar neutrino detector to the community of 
	solar neutrino experts. GENIUS will allow real time detection 
	of low-energy solar neutrinos with a threshold of 19 keV.
	   For the further potential of GENIUS for other beyond SM physics, such 
	as SUSY, compositeness, leptoquarks, 
	   violation of Lorentz invariance and equivalence principle, 
	   etc we refer to% 
\cite{KK-SprTracts00,KK-InJModPh98,KK60Y,KK-WEIN98,KK-Neutr98}.

%%%%%%%%%%%%%%%%%%%%%%%%% End Introduction %%%%%%%%%%%%%%%%%%%%%%%%%%

%%%%%%%%%%%%%%%%%%%%%%%%%% Section 1 %%%%%%%%%%%%%%%%%%%%%%%%%% 

%%%%%%%%% Fig. 1 %%%%%%%%%%%%%%%%%%%%%%%%%%%%%%%%%%%%%%%%
\begin{figure}[b]
\begin{picture}(60,140) 
\put(-90,210){\includegraphics{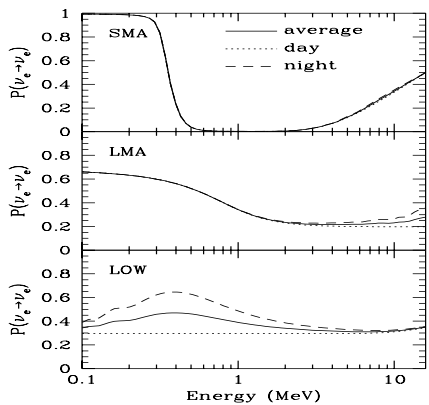}}
\end{picture}
\caption{%Figure 1. 
	Survival probabilities for an electron neutrino created in the 
	Sun for the three MSW solutions (from%
\protect\cite{Bach97}). 
	SMA, LMA, LOW stand for the small mixing angle, the large mixing 
	angle and the low $\Delta m^2$ MSW solutions.  
\label{fig:Bachal}
}
\end{figure}

%%%%%%%%% Fig. 1 %%%%%%%%%%%%%%%%%%%%%%%%%%%%%%%%%%%%%%%%

%%%%%%%%% Fig. 2 %%%%%%%%%%%%%%%%%%%%%%%%%%%%%%%%%%%%%%%%

\begin{figure} % Figure 2
\centering{
\includegraphics*[scale=0.37, angle=-90]
{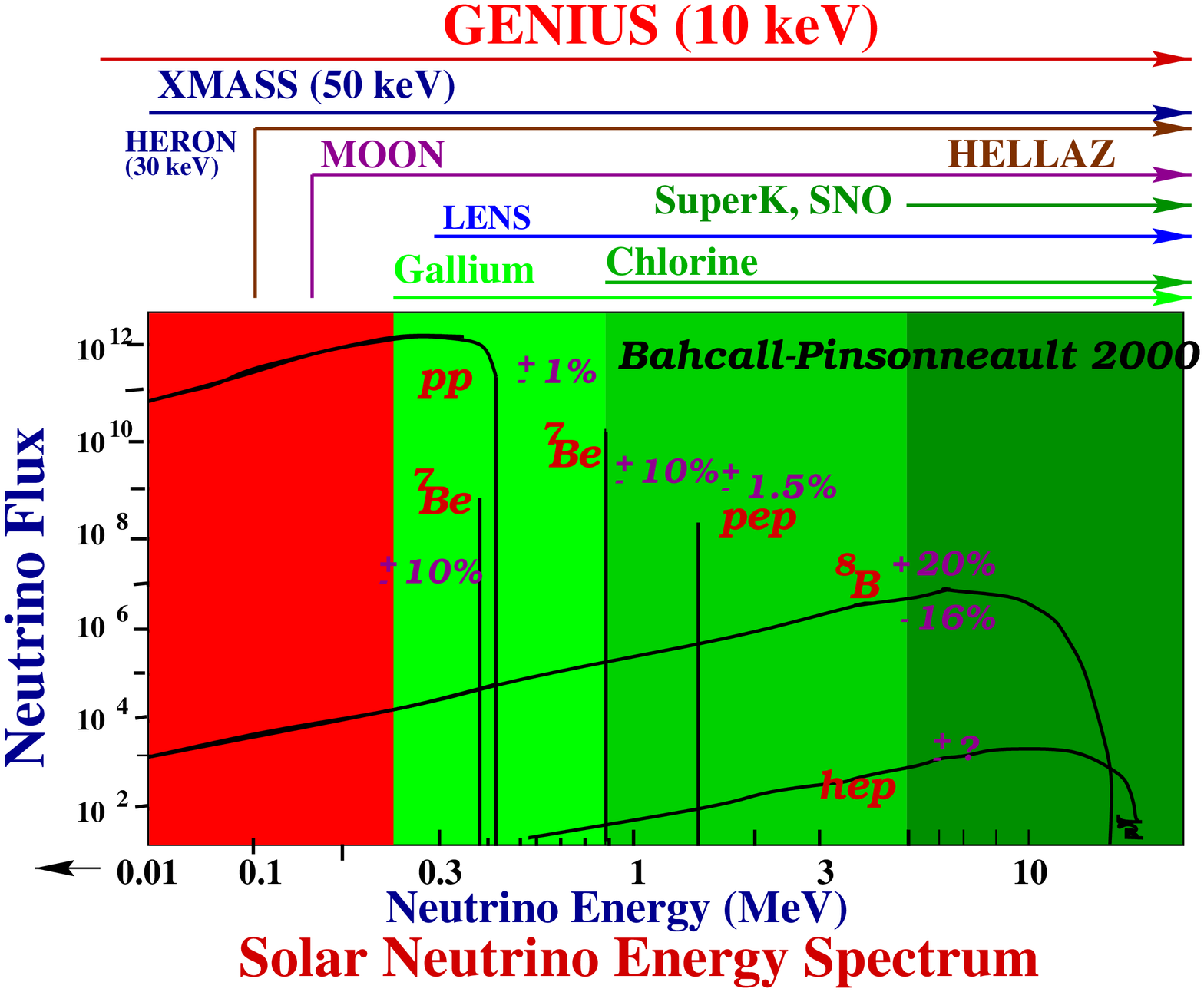}}
\caption[]{%Figure 12. 
       The sensitivity (thresholds) of different running and projected 
       solar neutrino detectors (see% 
\protect\cite{HomP-Bach} and HEIDELBERG NON-ACCELERATOR PARTICLE PHYSICS 
	GROUP home-page: http://www.mpi-hd.mpg.de/non\_acc/).
\label{fig:sol-neutr-Bach}}
%\end{figure}

%%%%%%%%% end Fig. 2 %%%%%%%%%%%%%%%%%%%%%%%%%%%%%%%%%%%%%%%%

%%%%%%%%%%%%%%%%%%%%%%%% Fig.3  %%%%%%%%%%%%%%%%%%%%%%%%%%%%%%%%%%%%%%%%

%\begin{figure} % Figure 3
\begin{center}
\includegraphics[width=.390\textwidth]{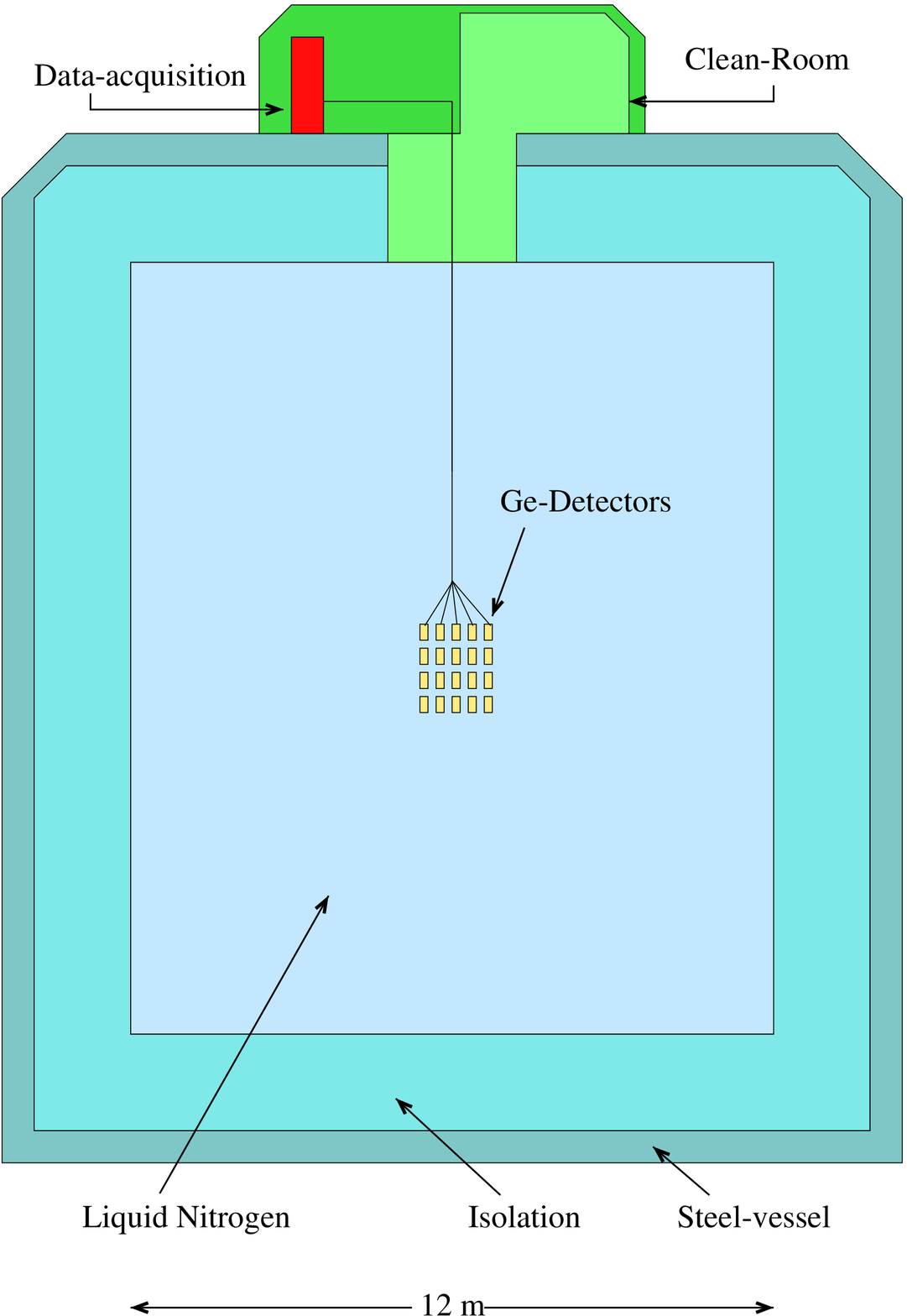} 
\end{center}
\caption[]{\label{genius_scheme}
	Schematic view of the GENIUS project. An array 
	of 100~kg of natural HPGe detectors for the WIMP dark matter search 
	(first step) or between 0.1 and 10 tons of
	enriched $^{76}$Ge for the double beta decay search (final setup) 
	is hanging on a support
	structure in the middle of the tank immersed in liquid nitrogen. 
	The size of the nitrogen shield would be 12 meters in diameter 
	at least. On top of the tank a special low-level clean room and 
	the room for the electronics and data 
	acquisition will be placed. }
\end{figure}

%%%%%%%%%%%%%%%%%%%%%%%% Fig. 3 %%%%%%%%%%%%%%%%%%%%%%%%%%%%%%%%%%%%%%%%

%%%%%%%%%%%%%%%%%%%%%%%%% sect. 1 %%%%%%%%%%%%%%%%%%%%

\section{GENIUS and Low-Energy Solar Neutrinos}

	To solve the solar neutrino problem, a measurement of the solar 
	low-energy spectrum in real time is required. The reasons are: 
	The prediction of the pp neutrino flux is almost solar-model 
	independent and strongly constrained by the solar luminosity and 
	helioseismological measurements. The spectral distortion by 
	oscillations allows to differentiate between different 
	oscillation solutions. 
	It is obvious from 
(Fig.~\ref{fig:Bachal})% 
\cite{Bach97}
	that a threshold below 300 keV is preferable, 
	i.e. lower than achieved by GALLEX and SAGE and also lower than those  
	of some future experiments.

	GALLEX and SAGE measure $pp +\, ^7{\rm Be} +\, ^8{\rm B}$ 
	neutrinos (60 + 30 + 10\%) down to 0.24 MeV, 
	the Chlorine experiment measured $^7{\rm Be} +\, ^8{\rm B}$ 
		neutrinos (80\% $^8$B) above $E_\nu= 0.817$~MeV, 
		all without spectral, time and direction information. 
		No experiment has separately measured the 
		$pp$ and $^7$Be neutrinos and no experiment has 
		measured the {\em full}\ $pp$ $\nu$ flux. 
		BOREXINO plans to measure $^7$Be neutrinos in real time, 
		the access to $pp$ neutrinos being limited by $^{14}$C 
		contamination (the usual problem of organic scintillators). 
		GENIUS which has been proposed for solar $\nu$ detection 
	in 1999% 
%\cite{Kla98d}
\cite{BKK-SolN}
	, could be the first detector measuring 
		the {\em full}\ $pp$ (and $^7$Be) 
		neutrino flux in real time (Fig.~\ref{fig:sol-neutr-Bach}).

	The main idea of GENIUS, originally proposed for double beta and dark 
	matter search%
\cite{KK-Bey97,KK-J-PhysG98,KK-Neutr98,KK-WEIN98,KK-InJModPh98,KK-SprTracts00}
  is to achieve an extremely low radiactive background 
	(factor of $>$  1000 smaller than in the HEIDELBERG-MOSCOW 
	expriment) 
	by using 'naked' detectors in liquid nitrogen 
(Fig.~\ref{genius_scheme}).

	While for double beta decay search a detector mass of 0.1---10 tons 
	of {\it enriched} $^{76}{Ge}$ is foreseen, and for cold 
	dark matter search 100 kg of {\it natural} Ge detectors 
	are sufficient, GENIUS as a solar neutrino detector would contain 
	1-10 tons of enriched $^{70}{Ge}$ or $^{73}{Ge}$.

	That Ge detectors in liquid nitrogen operate excellently, has been 
	demonstrated in the Heidelberg low-level laboratory% 
\cite{Hel97,Bau98}
	and the overall feasibility of the project has been shown in% 
\cite{GEN-prop}.

	The potential of GENIUS to measure the spectrum of low-energy solar 
	neutrinos in real time has been studied by% 
\cite{BKK-SolN,GEN-prop}. 
	The detection reaction is elastic neutrino-electon scattering 
	$\nu ~+~ e^- \longrightarrow~ \nu~ +~e^-$. In $\nu$--e scattering 
	experiments relatively higher statistics is achieved then in 
	absolution experiments (see Table 3), but no conicidence 
	information is available as in some of the absorption experiments.

	The maximum electron recoil energy is 261 keV for the pp neutrinos 
	and 665 keV for the $^{7}{Be}$ neutrinos% 
\cite{Bach89}. 
	The recoil electrons can be detected through their ionization 
	in a HP Ge detector with an energy resolution of 0.3$\%$. GENIUS 
	can measure only (like BOREXINO, and others) but with much better 
	energy resolution) the energy distribution of the recoiling electrons, 
	and not directly determine the energy of the incoming neutrinos. 
	The dominant part of the signal in GENIUS is produced by $pp$ 
	neutrinos (66$\%$) and $^{7}{Be}$ neutrinos (33$\%$). The detection 
	rates for the $pp$ and $^{7}{Be}$ fluxes are according to the 
	Standard Solar Model% 
\cite{BacBasPins98}
	$R_{pp}$ = 35 SNU = 1.8  events/day ton~ (18 events/day 10 tons) 
	and~ $R_{^{7}{Be}}$~ = 13 SNU~ = 0.6 events/day ton~ 
	(6 events/day 10 tons)~(1 SNU = ${10}^{-36}$/s target atom). 

	The expected total number of events, according to the SSM, and for 
	full $\nu_e \longrightarrow~ \nu_\mu$ conversion, are given  
	in Table 1.

%%%%%%%%%%%%%%%%%%%%%%% Table 1 %%%%%%%%%%%%%%%%%%%%%%%%%%%%%%%%%%%%%%
\begin{table}
\begin{center}
\newcommand{\m}{\hphantom{$-$}}
\newcommand{\cc}[1]{\multicolumn{1}{c}{#1}}
\renewcommand{\tabcolsep}{.95 pc} % enlarge column spacing
\renewcommand{\arraystretch}{.95} % enlarge line spacing
\begin{tabular}{|l|c|c|}
\hline
% & & \\
\vphantom{\Large I} & Events/day & Events/day  \\
& 11-665 keV& 11-665 keV\\
&  (1 ton) & (10 tons)\\
\hline
\hline
\vphantom{\Large I}
{\bf SSM}: & {\sf 2.4}   & {\sf 24.} \\
$pp$: 		 & {\it 1.8 (35 SNU)}  & {\it 18.} \\
$^7{Be}$: 	 &  {\it 0.6 (13 SNU)} & {\it 6.} \\
\hline
\vphantom{\Large I}
{\bf Full} $\nu_e \rightarrow \nu_{\mu}$ conversion & {\sf 0.62} & {\sf 6.2}\\
\hline
\vphantom{\Large I}
{\bf Background} & {\sf 0.6} & {\sf 6.}\\
\hline
\hline
\end{tabular}
\caption{Neutrino signal rates in GENIUS for 1 ton (10 tons) of 
Germanium.}
\label{rates}
\end{center}
\end{table}

%%%%%%%%%%%%%%%%%%%%%%%%%%%%%% end Table 1 %%%%%%%%%%%%%%%%%%%%%%%%%%%%%%%

\begin{figure}[h]
\centering{
\includegraphics*[scale=0.45]
{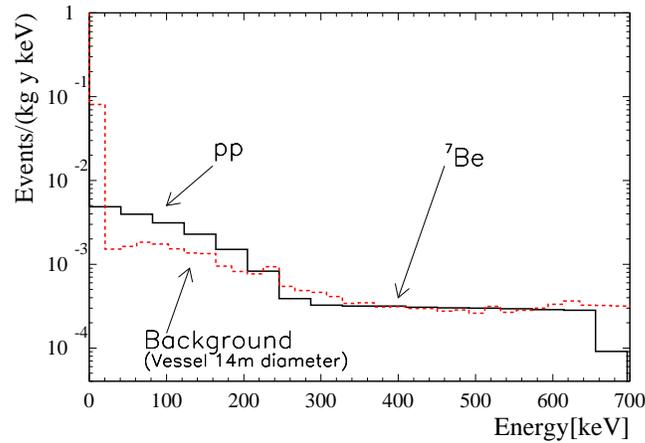}}

%\vspace{-.5cm}
\caption[]{%Figure 4. 
       Simulated spectrum of low-energy solar neutrinos 
       (according to SSM) for the GENIUS detector 
       (1 tonne of natural or enriched Ge) (from% 
\protect\cite{KK01}).
\label{fig:pp_7be-new}}
\end{figure}

	The expected spectrum of the low-energy signal in the SSM is 
	shown in 
Fig. ~\ref{fig:pp_7be-new}, 
	together with the total expected background 
	(which will be discussed below).

%%%%%%%%%%%%%%%%%%%%%%%%%%%%% begin. fig. 4  %%%%%%%%%%%%%%%
\vspace{-.5cm}
%\begin{figure}[h]
%\begin{picture}(-150,100) 
%\put(-50,-150){\special{PSfile=pp_7be-new.ps hscale=20 vscale=20 }}
%\end{picture}
%
%%%%%%%%%%%%%%%%%%%%%%%%%%%%%% end fig. 4  %%%%%%%%%%%%%%%

	Due to the excellent energy resolution of the detectors and the 
	difference in the cross 
	section of electron and muon neutrinos, a comparison of the energy 
	spectrum of recoiling electrons with the theoretical prediction 
	for various $\nu$ oscillation scenarios can be made to the following 
	extent. LMA and SMA MSW solutions can be differentiated only to 
	a limited degree. After the unfavouring of the SMA solution by 
	Superkamiokande, it may, however, now be more important to 
	differentiate between the LMA and the LOW solution. Here 
	due to its relatively high counting rate, GENIUS will be able to test  
	in particular the LOW solution of the solar $\nu$ problem by the 
	expected day/night variation of the flux 
(see Fig.~\ref{fig:Bachal}).

	If the signal to background ratio S/B will be greater than 1, then 
	the $pp$ and $^{7}{Be}$ fluxes can be measured by spectroscopic 
	techniques alone. If~~ S/B$<$1, one can make use of the seasonal 
	variation of the solar flux (7$\%$ from maximum to minimum) related 
	with the eccentricity of the Earth's orbit.

 %%%%%%%%%%%%%%%%%%%%%%%%%%%%%%%%%%%% fig. 5 %%%%%%%%%%%%%%%%%%%%%%
\vspace{-1.cm}
\begin{figure}[h]
\centering{
\includegraphics*[scale=0.41, angle=-90]{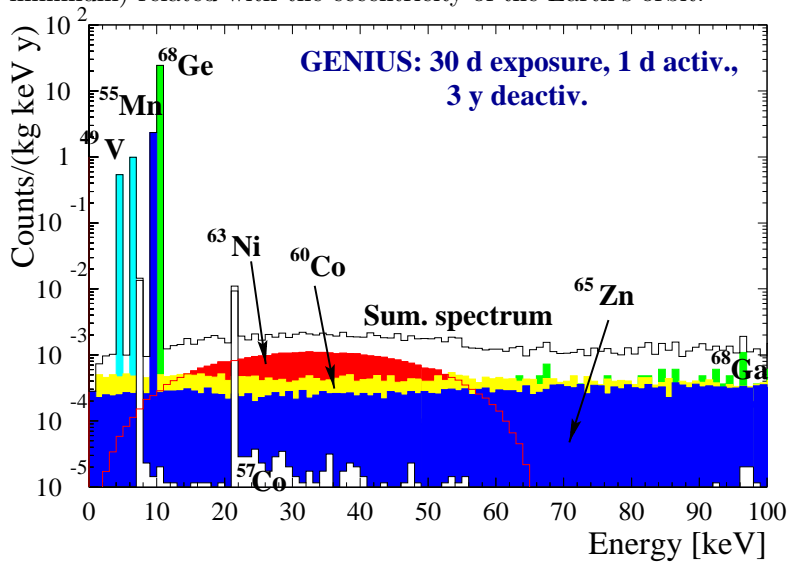}
}
%\figurebox{20pc}{15pc}
%\epsfbox{Cosmo-1d-3y.eps}
\vspace{-0.5cm}
\caption{%Figure 5. 
       Simulated cosmogenic background during detector production. 
       Assumptions: 30 days exposure of material before processing, 
       1 d activation after zone refining, 3 y deactivation 
       underground (neglecting tritium production) (see% 
\protect\cite{KK01,KK-NANPino00}).
%KK-LowNu2
\label{fig:Cosmo-1d-3y}}
\end{figure}

 %%%%%%%%%%%%%%%%%%%%%%%%%%%%%%%%%%%% end fig. 5 %%%%%%%%%%%%%%%%%%%%%%

\begin{table}[htb]
\begin{center}
\newcommand{\m}{\hphantom{$-$}}
\newcommand{\cc}[1]{\multicolumn{1}{c}{#1}}
\renewcommand{\tabcolsep}{.05 pc} % enlarge column spacing
\renewcommand{\arraystretch}{.95} % enlarge line spacing
\begin{tabular}{|l|c|c|c|}
\hline
{\sf Source} & {\sf Component} & {\sf Assumption} 
& {\sf [events/(kg y keV)]} \\
         &    &   & {\sf 11--260 keV}\\
\hline
\vphantom{\Large I} {\it LNitrogen}   & $^{238}$U, $^{232}$Th, $^{40}$K   
 & 3.5,  4.4,  10~$\times$~10$^{-16}$g/g &  3.6$\times$10$^{-4}$ \\
{\it contamination}           & $^{222}$Rn  & 0.5 $\mu$Bq/m$^3$  
&  2.5$\times$10$^{-5}$ \\
\hline
\vphantom{\Large I} {\it Steel vessel}   & U/Th        & 10$^{-8}$g/g 
&  4.5$\times$10$^{-5}$ \\
\hline
\vphantom{\Large I} {\it Holder system}  & U/Th        
& 10$^{-13}$g/g; 13g/det. 
&  8$\times$10$^{-5}$ \\
\hline
\vphantom{\Large I} {\it Surrounding}    & Gammas      
&  GS flux; tank: 13$\times$13m 
&  9$\times$10$^{-4}$\\
               & Neutrons    & GS flux  &   3$\times$10$^{-4}$\\
               & Muon shower & GS flux; muon veto 96$\%$  
&   7.2$\times$10$^{-6}$ \\
               & $\mu$ $\rightarrow$ n ($^{71}{Ge}$)      
& 230 capt. in nat. Ge/y  &   5$\times$10$^{-4}$ \\
%               & $\mu$ $\rightarrow$ capture & $<<$1$\times$10$^{-4}$ \\ 
\hline
{\it Cosmogenic}     & $^{54}$Mn, $^{57}$Co, $^{60}$Co,  
& 1d activ., 5y deactiv.  
&  {\Large $\}$} 8$\times10^{-4}$\\
 & $^{63}$Ni, $^{65}$Zn, $^{68}$Ge &  &  \\ 
\hline

{\it Total}          &   &  &  {\bf 3$\times$10$^{-3}$}  \\
\hline 
\end{tabular}
\end{center}
\caption{Simulated background sources together with the made assumptions 
	and the resulting event rates in GENIUS for the low-energy region 
	of the spectrum (from%
\protect\cite{BKK-SolN}).}
\label{backlist}
\end{table}

%%%%%%%%%%%%%%%%%%%%%%%%%%%%% Section 2 %%%%%%%%%%%%%%%%%%%%%%%%%%
\vspace{-0.5cm}
\section{Background Requirements}

	The average neutrino-induced signal is 3 $\times~{10}^{-3}$ 
	events/kg y keV in the energy region from 0 to 260 keV.  
	Thus, to measure the low-energy solar $\nu$ flux with a signal to 
	background ratio of 3:1, the required background rate is 
	about 1 $\times~ {10}^{-3}$ events/kg y keV in this energy range. 
	This is about a factor of 10 smaller than what is required for 
	the application of GENIUS for cold dark matter search. This can 
	be achieved if the liquid nitrogen shielding is increased to at 
	least 13 m in diameter and production of the Ge detectors is 
	performed underground. Table 2 shows the result of 
	corresponding Monte Carlo simulations (see% 
\cite{BKK-SolN}).
	Regarding the radiactivity of liquid nitrogen, the values reached 
	at present by BOREXINO for their liquid scintillator would be 
	sufficient. The external background from gamma rays from the Gran 
	Sasso laboratory could be further reduced by increasing the tank 
	beyond 13 m, or by further outside shielding.

%%%%%%%%%%%%%%%%%%%%%%%%%%%%%%% Fig. 6 %%%%%%%%%%%%%%%%%%%%%%%%%%%% 	
\begin{figure}[t]
\vspace{-0.7cm}
\centering{
\includegraphics*[scale=0.41, angle=-90]
{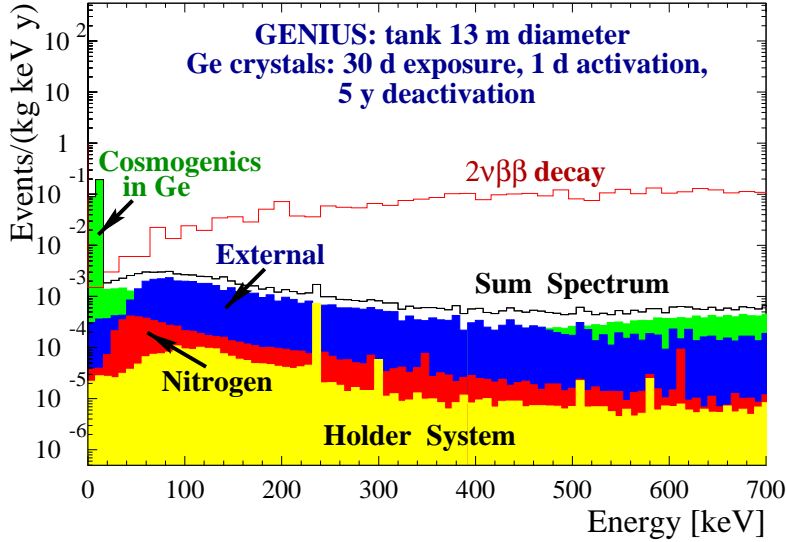}}
\vspace{-0.3cm}
\caption{%Figure 6. 
       Total background in a 13 m liquid nitrogen tank for detectors 
       produced as described in Fig. \ref{fig:Cosmo-1d-3y} 
	(tritium neglected) (see 
\protect\cite{KK01,KK-NANPino00}).
\label{fig:DBB-Spectr-13m-1d-5y-nonsat}}
\end{figure}

%%%%%%%%%%%%%%%%%%%%%%%%%%%%%%% end Fig. 6 %%%%%%%%%%%%%%%%%%%%%%%%%%%%

	Most attention has to be given to the cosmogenic activation of the 
	Ge crystals at the Earth surface. In 
Table 2  
	an overground 
	activation of only 1 day, and 5 years of deactivation have been 
	assumed. Figure 
\ref{fig:Cosmo-1d-3y} 
	shows the simulated cosmogenic background during detector 
	production, assuming 30 days of exposure of the material to cosmic 
	rays before detector production, one day of activation after zone 
	refining, and three years of deactivation underground 
	(neglecting tririum production, which would set a lower limit 
	(threshold) of the detector at $\sim$ 19 keV see%
\cite{KK-Maj-idm00}).

Fig. ~\ref{fig:DBB-Spectr-13m-1d-5y-nonsat}
	shows the {\it total} background in a 13 m liquid nitrogen tank 
	for detectors produced as described above, but with a deactivation 
	time of 5 years. Obviously the optimum solution would be to produce 
	the detectors in an underground facility. 
	
	Another source of background not included in the above discussion  
	is 2$\nu\beta\beta$ decay of $^{76}{Ge}$, which is contained in 
	{\it natural} Ge with 7.8$\%$. The expected rate for a detector of 
	natural Ge is shown in 
Fig. \ref{fig:DBB-Spectr-13m-1d-5y-nonsat}.
	The signal to 2$\nu\beta\beta$ background ration for the $pp$ 
	neutrino ranges from 3:1 to $\sim$ 1:100, and for the $^7{Be}$ 
	neutrinos is 1:100 to 1:1000. Using enriched 
	$^{70}{Ge}$ or $^{73}{Ge}$ ($>$85$\%$) 
	as detector material, the abundance of the $\beta\beta$ emitter 
	can be reduced up to a factor of 1500. In this case the $pp$-signal 
	is not disturbed by 2$\nu\beta\beta$ decay. 
	The $^7{Be}$ signal to $\beta\beta$ signal ratio will be about 1:1. 
	It can easily be extracted with the help of the seasonal modulation 
	of the flux originating from the earth's excentricity. 
	(The DAMA signal is 
	successfully extracted at a signal to background ratio of 1:10 
	by modulation).  
%	The $pp$ signal 
%	still can be extracted with the help of the seasonal modulation 
%	of the fluse from the Earth's excentricity. (The DAMA signal is 
%	successfully extracted at a signal to background ratio of 1:10 
%	by modulation). The $pp$ signal can be extracted easily also 
%	in the case of a natural Ge detector. 

%%%%%%%%%%%%%%%%%%%%%%%%%%%%% gorizontal Beg. Table 3 %%%%%%%%%%%%%%%%%%

\begin{table*}[h]
\caption{Some key numbers of running and future solar neutrino 
       experiments. 
%(see also \protect\cite{KK-LowNu2}).
}
\label{table:1}
\newcommand{\m}{\hphantom{$-$}}
\newcommand{\cc}[1]{\multicolumn{1}{c}{#1}}
\renewcommand{\tabcolsep}{.3 pc} % enlarge column spacing
\renewcommand{\arraystretch}{.65} % enlarge line spacing
{\footnotesize
%{\normalsize
{  
\begin{tabular}[!h]{|c|c|c|c|c|c|}
\hline
\hline
	& &   &   &   &          \\
	& Threshold  	& Resolution 	& Mass 	  	&  Reactions	
& Sensitivity    \\
	& (MeV) 		& (keV) 	& (tonnes) 	
&  		   	& ($\%$/ev/day )     \\
\cline{6-6}
\vphantom{\large I}
	&   &    &   &      
&   $^7$Be,\hspace{0.2cm}$^8$B, \hspace{0.2cm}$pp$\\
\hline
\hline
\vphantom{\large I} 
\underline{Water} 		& &   &   & Cerenkov  &    \\
SUPERK		&  5.  			&  	  		
& 50.000	&  $\gamma$-		   
&  $^8$B: 100$\%$~/~20-30$d^{-1}$   \\
(scatter.)	& 			&   
&  		&  rays 		    	&    \\
\hline
\vphantom{\large I} 
\underline{$^{37}$Cl} 	& &   &   &   &    \\
Davis		&  .817   		&  	  		
&  615.  	& $\nu_e~+~^{37}{Ar}\to $	   	
&   $\underbrace{^7{Be}: 20\%;~ ^8B: 80\%}_{\Large {\sim 1~d^{-1}}}$     \\
Experiment	& 			&   
&  $C_2{Cl}_4$ 	& $^{37}{Cl}+e^-$	   	
&     \\ 
(absorpt.)	& 			&  	  		
& 		& 		  	
&       \\
\hline
\vphantom{\large I} 
\underline{$^{71}$Ga} 	& &   &   &      & \\
(GALLEX)	& 0.235 		&  	  		
& 30. + 	& $^{71}{Ga}+\nu_e\to $    	
&      \\
GNO, 		& 			&   
&  		& 			   		
&  $^7$Be: 30$\%$   \\
SAGE		& 			&   
&  57.		& $^{71}{Ge}+e^-$ 	   	
&  $\underbrace{^8B: 10\%;~~pp: 60\%}_{\sim ~1~d^{-1}}$     \\
(absorpt.)		& 			&  	  		
& 		& 			   	&     \\
\hline
\vphantom{\large I} 
 \underline{$D_2$O} 	& 1.4 (CC) 		&  	  		
&  1 000.	& $\nu_e+d\to$		 		
&       \\
SNO		& 2.2 (NC)		&  	  		
& 		& $ p+p+e$  		     	
&  $^8$B: 100$\%$     \\
(scatter.)		& elast. - 0		
&  	  		
& 		&  $\nu_x+d\to$		    &       \\
		& 			&  	  		
& 		& $n+p+\nu_x$		    		
&       \\
\hline
\vphantom{\large I} 
\underline{$^{176}$Yb} 	&    &   &   &   & \\
LENS		& 0.241  		& 100.	  		
& 10.-30. 	& $^{176}{Yb}+\nu_e\to$     	
& $^7$Be:  0.2~$d^{-1}$    \\
(absorpt.) 	& 0.301 		&  	  		
& (8$\%$) enr.  & $^{176}{Lu}~+~e$		    
&  $pp$: 0.3$~d^{-1}$    \\
\hline
\vphantom{\large I} 
 \underline{$^{76}$Ge} &    &   &   &   & \\	
GENIUS		& {\bf 0.011} 		& 0.3$\%$ at   		
&  1.-10.	& $\nu+e^-\to $ 	   	
& $^7$Be:  33$\%$~/~.6-6~$d^{-1}$    \\
 (scatter.)	&  			&  300 keV 		
&  		& $ \nu+e^-$		    
&  $pp$: 66$\%$~/~1.8-18~$d^{-1}$    \\
&    &   &   &   &   \\
\hline
\vphantom{\large I} 
\underline{Scintill.}	&    &   &   &   & \\
BOREXINO$\ast$		& {\bf $\sim$ 0.030} 		&   	
&  100.$\ast$   & $\nu+e^-\to $ 	   	  
&  $^7$Be:  60 $d^{-1}\ast$   \\
KAMLAND$\dag$		& {\bf 0.050 or	}	& 4$\%\ast$ 	  	
& 1000.$\dag$   & $\nu+e^-$		   	  
&       \\
HERON$\ddag$ 		& {\bf 0.352}			&  	 	
& 10.$\ddag$   	& 			    
& $^7$Be:  6 $d^{-1}\sharp$       \\
XMASS$\sharp$		& 			&   
& 10.$\sharp$  	&  			     
& $pp$: 14$d^{-1}\sharp$   \\   
\hline
\vphantom{\large I} 
\underline{Drift}	&    & &  &   &    \\	
\underline{Chambers:}	& &    &   &   &    \\
HELLAZ$\ast\ast$ 	& 0.100$\ast\ast$ 	& 5$\%$ $\ast\ast$	
& 2000 $m^3$$\ast\ast$   & $\nu+e^-\to$     	  
& $^7$Be:  4 $d^{-1}$$\ast\ast$    \\
	(scatter.)	& 			&  	
& 		& $\nu+e^-$  $\ast\ast$     	  
& $pp$: 7 $d^{-1}$$\ast\ast$        \\
	MOON $\dag\dag$	& 0.168 $\dag\dag$	& 7$\%$$\dag\dag$   
&  3.3$\dag\dag$ 	& $\nu+^{100}{Mo}\to$  		     	  
& $^7$Be: 0.4 $d^{-1}$$\dag\dag$     \\
	(absorpt.)	& 			&   
&   			 &  $^{100}{Tc}$ + e 		     	  
& $pp$: 1.1 $d^{-1}$$\dag\dag$     \\

& &   &   &   &          \\
\hline
\hline 
\end{tabular}\\[2pt]
}}
%\vspace{0.3cm}
\end{table*}

%%%%%%%%%%%%%%%%%%%%%%%%%%%%%%%%%%%%%%end TABLE 3 %%%%%%%%%%%%%%%%%%%%%%%%
\vspace{-0.3cm}
\section{Comparison with other Detector Plans}

	 GENIUS will allow to look for the $pp$ and $^7$Be 
      solar neutrinos by elastic neutrino-electron scattering with a threshold
%      threshold of 11 keV or at most 19 keV 
%      (limit of possible tritium background)
(Fig.~\ref{fig:sol-neutr-Bach})
%~\ref{fig:pp_7be-new}) 
	which would be the lowest threshold among other proposals to detect 
	$pp$ neutrinos, such as 
	HERON% 
\cite{LowNu2}, 
	HELLAZ% 
\cite{LowNu2}, 
	NEON% 
\cite{LowNu2}, 
	LENS% 
\cite{Fuj00,LowNu2}, 
	MOON% 
\cite{LowNu2}, 
	XMASS% 
\cite{XMAXX00,LowNu2}.

	The counting rate of GENIUS (10 ton) would be 6 events per day 
	for $pp$ and 18 per day for $^7$Be neutrinos, i.e. similar to 
	BOREXINO, but by a factor of 30 to 60 larger than a 20 ton LENS 
	detector and a factor of 10 larger than the MOON detector (see 
Table 3).

	The good energy resolution for detecting the recoiling electrons 
	would allow for the first time to measure the 1.3 keV predicted 
	shift of the average energy of the beryllium neutrino line. 
	This shift is a direct measure of the central temperature of the Sun% 
\cite{Bach93}.

%%%%%%%%%%%%%%%%%%%%%%%%%%%%% Section 4 %%%%%%%%%%%%%%%%%%%%%%%%%%%%% 
\vspace{-0.3cm}
\section{GENIUS and the Neutrino Mass from Double Beta}

	Another -- more general -- contribution to neutrino physics 
	by GENIUS comes from its potential to investigate double beta decay. 
	It has been shown recently% 
\cite{KKPS,KKPS-01}
	that by double beta experiments it will be possible to test 
	almost all neutrino mass scenarios allowed by the present 
	neutrino oscillation experiments.

%%%%%%%%%%%%%%%%% begin. fig. 6 *************************
%\clearpage
\begin{figure}[b]
%\vspace{9pt}
\centering{
\includegraphics*[scale=0.42, angle=-90]
{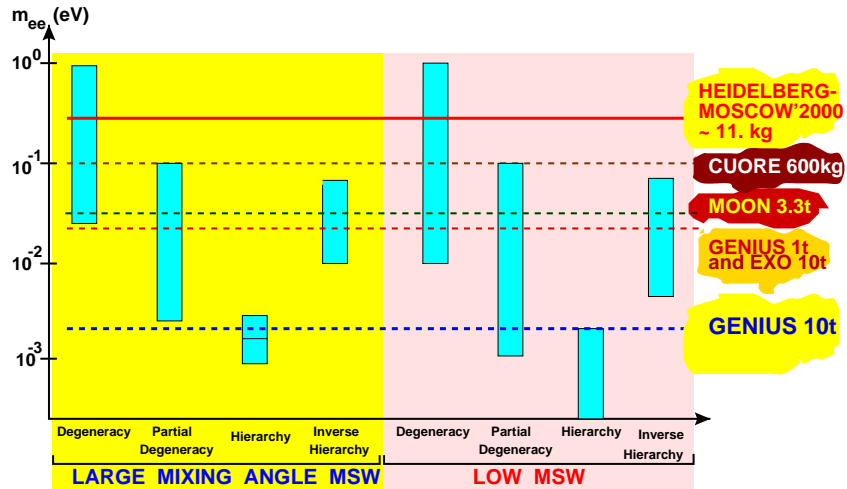}}
\caption[]{%Figure 6. 
       	Summary of values for $m_{ee} =\langle m \rangle$
	expected from neutrino oscillation experiments 
	(status NEUTRINO2000), in the two neutrino mass schemes
	presently favored by solar $\nu$ experiments.  
	For a more general analysis see% 
\cite{KKPS}. 
	The expectations are compared with the recent neutrino mass 
	limits obtained from the HEIDELBERG-MOSCOW experiment% 
\cite{KK-PhysRevD01,AnnRepGrSs00}
%\cite{KK01,AnnRepGrSs00}
	as well as the expected sensitivities for the CUORE,  
%\cite{CUORE98},  
	MOON,  
%\cite{Ej00}, 
	EXO  
%\cite{EXO00} 
	proposals and the 1 ton and 10 ton proposal of GENIUS% 
\cite{KK-Bey97,GEN-prop} 
	(for details see%
\cite{KK60Y,KK-NOON00}).
\label{fig:Jahr00-Sum-difSchemNeutr}}
\end{figure}

%%%%%%%%%%%%%%%%% end fig. 6 *************************

	This is shown in 
Fig.~\ref{fig:Jahr00-Sum-difSchemNeutr}
	for the Large Mixing Angle and LOW MSW solutions, presently favored 
	by solar neutrino oscillation experiments. Nature seems to be 
	generous to us, that with an increase of the sensitivity for 
	the effective neutrino mass 
$\langle m \rangle$ 
	- the observable of double beta 
	decay --- by two orders of magnitude compared to the present best 
	limit (HEIDELBERG-MOSCOW experiment%
\cite{KK-PhysRevD01}
	) almost the {\it full range} of expectations 
	from oscillation experiments can be covered. This is possible, 
	however, only by exploiting this method to obtain information 
	on the neutrino mass to the ultimate limit. This can be done 
	by the GENIUS project 
(for details see%
\cite{KKPS,KKPS-01,KK60Y}).

%%%%%%%%%%%%%%%%%%%%%%%%%%%%% Section 6 %%%%%%%%%%%%%%%%%%%%%%%%%%%%% Section 

\section{GENIUS and Cold Dark Matter Search}
	
	GENIUS is also the only of the new projects under discussion 
(see Tabl. 4) 
%(fig. ~\ref{fig:NewProj1-catMaj}
	which simultaneously with its potential for real-time detection 
	of low-energy neutrinos, and for double beta decay, has a huge 
	potential for cold dark matter search.

%%%%%%%%%%%%%%%%%% tabl. 2 %%%%%%%%%%%%%%%%%%%%%%%%%%%
\begin{table*}[h]
\caption{Some of the new projects under discussion for future double beta 
	decay experiments (see ref.%
\protect\cite{KK60Y}).}
\label{tableA}
\newcommand{\m}{\hphantom{$-$}}
\newcommand{\cc}[1]{\multicolumn{1}{c}{#1}}
\renewcommand{\tabcolsep}{.19 pc} % enlarge column spacing
\renewcommand{\arraystretch}{.15} % enlarge line spacing
\begin{tabular}[!h]{|c|c|c|c|c|}
\hline
\hline

\multicolumn{5}{|c|}{}\\
\multicolumn{5}{|c|}{\Large NEW~~~  PROJECTS}\\
\multicolumn{5}{|c|}{}\\
\hline
 &  &  &  & \\
\vphantom{\large I}
 & BACKGROUND & MASS & POTENTIAL & POTENTIAL \\
 &  &  &  & \\
 & REDUCTION & INCREASE & FOR DARK & FOR SOLAR\\
 &  &  & MATTER & ${\nu}^{'}$ s\\
\hline 
\hline
&  &  &  & \\
 {\sf GENIUS} 	& {\Large\bf +} 	& {\Large\bf +} 	
& {\Large\bf +} & {\Large\bf +$^{*)}$	}\\
\hline
&  &  &  & \\
 {\sf XMASS} 		&  {\Large\bf +} 	& {\Large\bf +}  	
& {\Large\bf +} & {\Large\bf +$^{*)}$	}	\\
\hline
&  &  &  & \\
 {\sf CUORE}  	& {\Large\bf (+)} 	& {\Large\bf + } 	
& {\Large\bf $-$} & {\Large\bf $-$	}	\\
\hline
&  &  &  & \\
 {\sf MOON}  		& {\Large\bf (+) }	& {\Large\bf + } 	
& {\Large\bf $-$} & {\Large\bf +}		\\
\hline
&  &  &  & \\
 {\sf EXO} 		&  {\Large\bf +} 	& {\Large\bf +}  	
& {\Large\bf $-$} & {\Large\bf $-$	}	\\
\hline
&  &  &  & \\
 {\sf MAJORANA}	& {\Large\bf $-$ }	& {\Large\bf + } 	
& {\Large\bf $-$} & {\Large\bf $-$	}	\\
\hline
\multicolumn{5}{|c|}{}\\
\multicolumn{5}{|c|}{\Large\sf *) real time measurement of pp neutrinos}\\
\multicolumn{5}{|c|}{\Large\sf with threshold of 10--50 keV (!!)}\\
\multicolumn{5}{|c|}{}\\
\hline
\hline
\end{tabular}\\[1pt]
%\vspace{0.3cm}
\end{table*}

%%%%%%%%%%%%%%%%%% end tabl. ? %%%%%%%%%%%%%%%%%%%%%%%%%%%

%%%%%%%%%%%%%%%%%%%%  fig. 9 ****************************

%\clearpage
\begin{figure}
\begin{picture}(100,145)
\centering{
\put(40,-140){\includegraphics{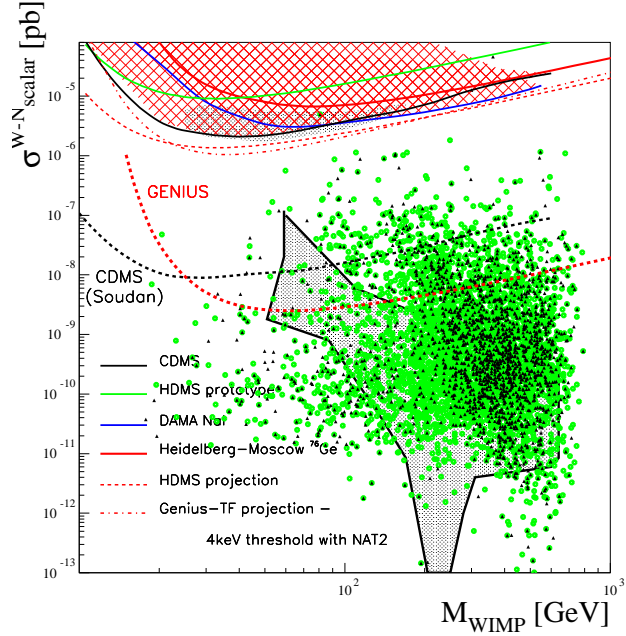}}}
\end{picture} 

\vspace{2.3cm}
\caption[]{%Figure 9. 
       WIMP-nucleon cross section limits in pb for scalar interactions as 
       function of the WIMP mass in GeV. 
       Shown are contour lines of present experimental limits (solid lines) 
       and of projected experiments (dashed lines). 
       Also shown is the region of evidence published by DAMA. 
       The theoretical expectations are shown by a scatter plot (from% 
\cite{BedKK00}) and by the grey region (from% 
\cite{EllOliv-DM00}). 
	{\em Only}\ GENIUS will be able to probe the shown range 
       also by the signature from seasonal modulations.
\label{fig:Bedn-Wp2000}}
\end{figure}

%%%%%%%%%%%%%%%%%%%%%%%%  end fig. 9 ***************************

	Already now the HEIDELBERG-MOSCOW experiment is the most 
		sensitive Dark Matter experiment worldwide concerning 
		the raw data% 
\cite{KK-PRD63-00,HM98,Ram99,KK-dark00}. 
	GENIUS would already in a first step, with 100 kg of 
		{\it natural} Ge detectors, cover a significant part of the 
		MSSM parameter space for prediction of neutralinos 
		as cold dark matter 
(Fig.~\ref{fig:Bedn-Wp2000}) 
	(see, e.g.%
\cite{KKRam98}). For this purpose the background in the energy range 
		$< 100$~keV has to be reduced to 
		$10^{-2}$ events/(kg y keV), 
		which is possible if the detectors are produced 
		and handled on Earth surface under heavy shielding, 
		to reduce the cosmogenic background produced by 
		spallation through cosmic radiation (critical products are 
		tritium, $^{68}$Ge, $^{63}$Ni, ...) to a minimum. 
		For details we refer to% 
\cite{GEN-prop,DARK2000,KK-Maj-idm00}. 
Fig.~\ref{fig:pp_7be-new}
	shows that at this level solar neutrinos as source of background 
	are still negligible. 
Fig.~\ref{fig:Bedn-Wp2000} 
	shows together with the expected sensitivity of GENIUS, 
	for this background, predictions for neutralinos as dark matter by 
	two models, one basing on supergravity% 
\cite{EllOliv-DM00}, another starting from more 
     relaxed unification conditions% 
\cite{BedKK00,BedKK-01}.

	     The sensitivity of GENIUS for Dark Matter corresponds to 
	     that obtainable with a 1 km$^3$ AMANDA detector for 
	     {\it indirect} detection (neutrinos from annihilation 
	     of neutralinos captured at the Sun) (see% 
\cite{Eds99}). 
	Interestingly both experiments would probe different neutralino 
	compositions: GENIUS mainly gaugino-dominated neutralinos, 
	AMANDA mainly neutralinos with comparable gaugino and 
	Higgsino components (see Fig. 38 in% 
\cite{Eds99}). 
     It should be stressed that, together with DAMA, GENIUS will be 
     {\em the only}\ future Dark Matter experiment, which would be able to 
     positively identify a dark matter signal by the seasonal 
     modulation signature. 
     This {\it cannot} be achieved, for example, by the CDMS experiment.

%%%%%%%%%%%%%%%%%%%%%%%%%%%%%%%%%% Section 6 %%%%%%%%%%%%%%%%%%%%%%%%

\section{Conclusion}

	Concluding, GENIUS could measure 
	solar pp neutrinos in real time. The detector reaction would be 
	elastic neutrino-electron scattering, with a threshold of 11 or 
	at most 19 keV (limit of possible tritium background), which would 
	be the lowest threshold among other proposals, to detect $pp$ 
	neutrinos, such as 
	HERON% 
\cite{LowNu2}, 
	HELLAZ% 
\cite{LowNu2}, 
	NEON% 
\cite{LowNu2}, 
	LENS% 
\cite{Fuj00,LowNu2}, 
	MOON% 
\cite{LowNu2}, 
	XMASS% 
\cite{XMAXX00,LowNu2}.
	
	The counting rate of GENIUS (10 ton) of 6 events per day for $pp$ 
	and 18 per day for $^{7}{Be}$ neutrinos, would be similar to 
	BOREXINO, but by a factor of 30-60 larger than a 20 ton LENS 
	detector and a factor of 10 larger than the MOON datector 
(see Table 3).

	The good energy resolution for detecting the recoiling electrons 
	would allow for the first time to measure the 1.3 keV predicted 
	shift of the average energy of the beryllium neutrino line. 
	This shift is a direct measure of the central 
	temperature of the Sun% 
\cite{Bach93}.

	GENIUS would further allow -- in its main applications -- to probe, 
	by search for double beta decay, almost the full range of 
	expectations for the effective neutrino 
	mass in practically all neutrino mass scenarios, consistent with 
	present neutrino oscillation experiments. Finally it would be the 
	most sensitive cold dark matter detector, having the 
	potential to probe a significant part of the MSSM parameter space 
	for prediction of neutralinos as cold dark matter. It will be the 
	{\it only} future dark matter experiment (besides DAMA), which 
	would be able to positively by identify a dark matter signal by the 
	seasonal modulation signature.

%%%%%%%%%%%%%%%%%%%%%%%%%%%%%%%%%%%%THE BIBLIOGR>%%%%%%%%%%%%%%%%%


\begin{thebibliography}{99}

\bibitem{KKPS}%{1}
	H.V. Klapdor-Kleingrothaus, H. P\"as and A.Yu. Smirnov, 
        {\it Preprint:} {\it hep-ph/}{\bf 0003219}, (2000) and in 
	{\it Phys. Rev.} {\bf D} (2000).

\bibitem{KKPS-01}%{2}
	H.V. Klapdor-Kleingrothaus, H. P\"as and A.Yu. Smirnov, 
	in Proc. of DARK2000, Heidelberg, 10-15 July, 2000, 
	Germany, ed.  H.V. Klapdor-Kleingrothaus, 
	{\it Springer, Heidelberg} (2001) and 
	{\it Preprint:} {\it hep-ph/}{\bf 0103076} .

\bibitem{KK60Y}%{3}	
		H.V. Klapdor-Kleingrothaus, 
		{\sf "60 Years of Double Beta Decay"}, {\it World Scientific, 
		Singapore} (2001) 1253~p.

\bibitem{KKP-ComNucl}%{4}	
	H.V. Klapdor-Kleingrothaus, H. P\"as, 
		{\it Preprint: physics/}{\bf 0006024} and 
		{\it Comm. in Nucl. and 
		Part. Phys.} (2000).

\bibitem{Hel97}%{5}
	J. Hellmig and H.V. Klapdor--Kleingrothaus, 
	{\it Z. Phys.} {\bf A 359}, 351 (1997).

\bibitem{Bau98}%{6}
 	L.~Baudis, G.~Heusser, B.~Majorovits, Y.~Ramachers, H.~Strecker and 
 	H.V.~Klapdor--Kleingrothaus, {\it hep-ex/} {\bf 9811040} and 
	{\it Nucl. Instr. Meth.} {\bf A 426}, 425 (1999).


\bibitem{BKK-SolN}%{7}
	L. Baudis and H.V. Klapdor-Kleingrothaus, 
	{\it Eur. Phys. J.} {\bf A 5}, 441-443  (1999) and in 
	Proceedings of the 2nd Int. Conf. on Particle 
	Physics Beyond the Standard Model BEYOND'99, 
	Castle Ringberg, Germany, 6-12 June 1999, 
	edited by H.V. Klapdor-Kleingrothaus and I.V. Krivosheina, 
	{\it IOP Bristol}, 1023 - 1036 (2000).

\bibitem{KK01}%{8}	
		H.V. Klapdor-Kleingrothaus et al., 
		to be publ. 2001 
		and ${\it http://www.mpi-hd.mpg.de/non\_acc/}$

\bibitem{HM98}%{9}	
		HEIDELBERG-MOSCOW Collaboration, {\it Phys. Rev.} 
	{\bf D 59}, 022001 (1998).

\bibitem{KK-PRD63-00}%{10}
	L. Baudis, A. Dietz, B. Majorovits, F. Schwamm, H. Strecker 
	and H.V. Klapdor-Kleingrothaus, {\it Phys. Rev.} 
	{\bf D 63 }, 022001 (2000) and 
	{\it astro-ph/}{\bf 0008339}.

\bibitem{Ram99}%{11}
	Y. Ramachers for the CRESST Collaboration in Proc. of 
	XIth Rencontres de Blois, Frontiers of Matter, France, 
	June 27-July 3, 1999.

%%%%%%%%%% DARK2000 HEIDELBERG %%%%%%%%%%%%%%%%%

\bibitem{DARK2000}%{12} 
	H.V. Klapdor-Kleingrothaus et al. Proc. DARK2000, Heidelberg, 
	Germany, July 10-15, 2000, Ed.  H.V. Klapdor-Kleingrothaus, 
	{\it Springer, Heidelberg} (2001).

\bibitem{KK-dark00}%{13}
	H.V. Klapdor-Kleingrothaus et al. in 
	Proc. of Third International Conference on Dark Matter in 
	Astro and Particle Physics, DARK2000, Heidelberg, 
	Germany, July 10-15, 2000, {\it Springer, Heidelberg} (2001), 
	ed. H.V. Klapdor-Kleingrothaus 
	and {\it Preprint: hep-ph/}{\bf 0103082}.

\bibitem{KK-PhysRevD01}%{14}
	H.V. Klapdor-Kleingrothaus et al. in Proc. of Third International 
	Conference on Dark Matter in 
	Astro and Particle Physics, DARK2000, Heidelberg, 
	Germany, July 10-15, 2000, {\it Springer, Heidelberg} (2001), 
	ed. H.V. Klapdor-Kleingrothaus 
	and {\it Preprint: hep-ph/}{\bf 0103062}, subm. to {\it Phys. Rev.} 
	{\bf D} (2001).

%%%%%%%%%%%%%%%%% NANPino Conf %%%%%%%%%%%%%%%%%%%%%%%%%%%%%%%%

\bibitem{KK-NANPino00}%{15} 
	H.V. Klapdor-Kleingrothaus, 
	 In Proc. of International Workshop Non-Accelerator 
	New Physics in neutrino observations, NANPino-2000, Dubna, 
	July 19-22, 2000, ed. V. Bednyakov et al., 
	to be publ. in {\it Particles and Nuclei, Letters}, 
	issues {\bf 1/2} (2001) and {\it Preprint: hep-ph/}{\bf 0102319}.

%%%%%%%%%%%%%%%%%%%%%%% IDM2000 England %%%%%%%%%%%%%%%%%%%

\bibitem{KK-Maj-idm00}%{16}
	H.V. Klapdor-Kleingrothaus and B. Majorovits, 
	in Proc. of 3rd International Workshop on the Identification of 
	Dark Matter (IDM2000), York, England, 18-22 Sep 2000, 
	{\it World Scientific, Singapore } (2001)  
	and Preprint: {\it hep-ph/}{\bf 0103079}.

%%%%%%%%%%%%%%%% BEY97 %%%%%%%%%%%%%%%%%%%%%%%
 	
\bibitem{KK-Bey97}%{17}
	H.V. Klapdor-Kleingrothaus in Proceedings of BEYOND'97, 
	First International Conference on Particle Physics 
	Beyond the Standard Model, Castle Ringberg, Germany, 
	8-14 June 1997, 
     edited by H.V. Klapdor-Kleingrothaus and H. P\"as, 
	{\it IOP Bristol} 485-531 (1998)  
%	and {\it Int. J. Mod. Phys.} {\bf A 13} (1998) 3953, and 
%     {\it J. Phys.} {\bf G 24} (1998) 483-516.

%%%%%%%%%%%%%%%%%%%%% BEY99 %%%%%%%%%%%%%%%%%%%%%%%%

\bibitem{GEN-prop}%{18}
	H.V. Klapdor-Kleingrothaus et al. 
	{\it MPI-Report} {\bf MPI-H-V26-1999} and 
	{\it Preprint: hep-ph/}{\bf 9910205} and in 
	Proceedings of the 2nd Int. Conf. on Particle 
	Physics Beyond the Standard Model BEYOND'99, 
	Castle Ringberg, Germany, 6-12 June 1999, 
	edited by H.V. Klapdor-Kleingrothaus and I.V. Krivosheina, 
	{\it IOP Bristol}, 915 - 1014 (2000).

%%%%%%%%%%%%% NEUTRINO98 %%%%%%%%%%%%%
\bibitem{KK-Neutr98}%{19}
	H.V. Klapdor-Kleingrothaus, in Proc. of 18th Int. 
	Conf. on Neutrino Physics and Astrophysics (NEUTRINO 98), 
	Takayama, Japan, 4-9 Jun 1998, (eds) Y. Suzuki et al. 
	{\it Nucl. Phys. Proc. Suppl.} {\bf 77}, 357 - 368 (1999).   

%%%%%%%%%%%%%%%%%%%%5 WEIN98 %%%%%%%%%%%%%%%%%%%%%%%

\bibitem{KK-WEIN98}%{20}
	H.V. Klapdor-Kleingrothaus, in Proc. of WEIN'98, 
	"Physics Beyond the Standard Model", Proceedings of the Fifth Intern.  
	WEIN Conference, P. Herczeg, C.M. Hoffman and 
	H.V. Klapdor-Kleingrothaus (Editors), 
	{\it World Scientific, Singapore}, 275-311 (1999).

%%%%%%%%%%%%%%%%%%%%%%%%%%% Ann. Rep.; Gr-SS and MPI %%%%%%%%%%%%

\bibitem{AnnRepGrSs00}%{21}
	H.V. Klapdor-Kleingrothaus et al., 
	{\it Annual Report Gran Sasso 2000} (2001).

%%%%%%%%%%%%%%%%%%%%%%%%%%%%%%%%%%%%%% NOW 2000 %%%%%%%%%%%%%

\bibitem{KK-NOW00}%{22}
	H.V. Klapdor-Kleingrothaus, in Proc. of Int. Conference 
	NOW2000 - "Origins of Neutrino Oscillations", 
	{\it Nucl. Phys.} {\bf B} (2001) ed. G. Fogli and 
	{\it Preprint: hep-ph/}{\bf 0102277}, 
	{\it Preprint: hep-ph/}{\bf 0102276}. 

%%%%%%%%%%%%%%%%%%%%%%%%%% NOON 2000 and LowNu2000 %%%%%%%%%%%%%%%%%

\bibitem{KK-NOON00}%{23}
	H.V. Klapdor-Kleingrothaus, in Proc. of NOON2000, 
	International Workshop on "Neutrino Oscillations and Their Origin", 
	Tokyo, 6-18 Dec. 2000, {\it World Scientific, Singapore} (2001) 
	and {\it Preprint: hep-ph/}{\bf 0103074}.

\bibitem{LowNu2}%{24}
	Proc. Int. Workshop on Low Energy Solar Neutrinos, LowNu2, 
	December 4 and 5 (2000) Tokyo, Japan, 
	ed: Y. Suzuki, World Scientific, Singapore (2001), home page: 
	{\it http://www-sk.icrr.u-tokyo.ac.jp/neutlowe/2/transparency/
index.html}

\bibitem{XMAXX00}%{25}
	Y. Suzuki for the collaboration, {\it Preprint: hep-ph/}{\bf 0008296}.

\bibitem{Fuj00}%{26}
	M. Fujiwara et al., {\it Phys. Rev. Lett.} {\bf 85}, 4442-4445 (2000) 
	and {\it Preprint: nucl-ex/}{\bf 0006006}: 
	M. Bhattacharya et al. {\it Phys. Rev. Lett.} {\bf 85}, 4446-4449 
	(2000) and 
	{\it Preprint: nucl-ex/}{\bf 0006005}.

%%%%%%%%%%%%% Journ %%%%%%%%%%%%%%%%

\bibitem{KK-J-PhysG98}%{27}
	H.V. Klapdor-Kleingrothaus, J. Hellmig and M. Hirsch, 
	{\it J. Phys.} {\bf G 24}, 483 (1998).

\bibitem{KK-InJModPh98}%{28}
	H.V. Klapdor-Kleingrothaus, {\it Int. J. Mod. Phys.} {\bf A 13}, 3953  
	(1998).
%	and {\it J. Phys.} {\bf G 24} (1998) 483--516.

\bibitem{KK-SprTracts00}%{29}
	H.V. Klapdor-Kleingrothaus, {\it Springer Tracts in Modern Physics}, 
	{\bf 163}, 69-104 (2000), 
	{\it Springer-Verlag, Heidelberg, Germany} (2000).

%%%%%%%%%%%%%%%% Dark Matter %%%%%%%%%%%

\bibitem{KKRam98}%{30}
	H.V. Klapdor-Kleingrothaus and Y. Ramachers, 
	{\it Eur. Phys. J.} {\bf A 3}, 85-92 (1998).

\bibitem{BedKK00}%{31}
	V.A. Bednyakov and H.V. Klapdor-Kleingrothaus, 
	{\it Phys. Rev.} {\bf D 62} (2000) 043524/1-9 and {\it hep-ph/}
	{\bf 9908427}. 

\bibitem{BedKK-01}%{32}
	V.A. Bednyakov and H.V. Klapdor-Kleingrothaus, 
	{\it Preprint: hep-ph/}{\bf 0011233} (2000) in 
	press in {\it Phys. Rev.} {\bf D} (2001).

\bibitem{EllOliv-DM00}%{33}
	J. Ellis, A. Ferstl and K.A. Olive, {\it Phys. Lett.} {\bf B 481},  
	304--314 (2000) and {\it Preprint: hep-ph/}{\bf 0001005} 
	and {\it Preprint: hep-ph/}{\bf 0007113}.

\bibitem{Eds99}%{34}
	J. Edsj\"o, Neutralinos as dark matter - can we see them? 
	Seminar given in the theory group, Department of Physics, 
	Stockholm University, October 12, 1999, home page: 
	{\it http://www.physto.se/edsjo/}


%%%%%%%%%%%%%%% Solar Neutrinos Bahcall %%%%%%%%%

\bibitem{HomP-Bach}%{35}
	see: {\it http://www.sns.ias.edu.jnb/}

\bibitem{Bach89}%{36}
	J.N. Bahcall, {\it Neutrino Astrophysics}, Cambridge Univ. 
	Press (1989). 

\bibitem{Bach93}%{37}
	J.N. Bahcall, {\it Phys. Rev Lett.} {\bf  71}, 2369 (1993).

\bibitem{Bach97}%{38}
	J.N. Bahcall and P.I. Krastev, {\it Phys. Rev.} {\bf C 56}, 
	2839 (1997).

\bibitem{BacBasPins98}%{39}
	J.N. Bahcall, S. Basu and M. Pinsonneault, 
	{\it Phys. Lett.} {\bf B 433}, 1 (1998).


\end{thebibliography}
\end{document}